\documentclass[twocolumn,prl,aps,epsfig]{revtex4}
\usepackage{epsfig}
\def\beq{\begin{equation}}
\def\enq{\end{equation}}
\def\beqa{\begin{eqnarray}}
\def\enqa{\end{eqnarray}}
\def\MeV{\nobreak\,\mbox{MeV}}
\def\GeV{\nobreak\,\mbox{GeV}}

\def\qq{\lag\bar{q}q\rag}
\def\ss{\lag\bar{s}s\rag}
\def\mix{\lag\bar{q}g\si.Gq\rag}
\def\mixs{\lag\bar{s}g\si.Gs\rag}
\def\Gd{\lag g^2G^2\rag}
\def\G3{\lag g^3G^3\rag}

\def\La{\Lambda}

\def\rh{\rho}
\def\si{\sigma}

\def\al{\alpha}

\def\lb{\label}
\def\nn{\nonumber}
\newcommand{\rag}{\rangle}
\newcommand{\lag}{\langle}

\begin{document}

\title{\sc Disentangling two- and four-quark state pictures of the charmed 
scalar mesons}
\author{M.E. Bracco$^1$, A. Lozea$^1$, R.D. Matheus$^2$, F. S. Navarra$^2$ 
and M. Nielsen$^2$}
\affiliation{$^1$Instituto de F\'{\i}sica, Universidade do Estado do Rio de 
Janeiro, 
Rua S\~ao Francisco Xavier 524, 20550-900 Rio de Janeiro, RJ, Brazil//
$^2$Instituto de F\'{\i}sica, Universidade de S\~{a}o Paulo, 
C.P. 66318, 05389-970 S\~{a}o Paulo, SP, Brazil}

\begin{abstract}
We suggest that the recently observed charmed scalar mesons 
$D_0^{0}(2308)$ (BELLE) and $D_0^{0,+}(2405)$ (FOCUS) are considered as
different resonances. Using the QCD sum rule approach we investigate the 
possible four-quark structure of these mesons and also of the very narrow
$D_{sJ}^{+}(2317)$, firstly observed by BABAR. We use  diquak-antidiquark 
currents and work to the order of $m_s$ in full QCD, without relying on 
$1/m_c$ expansion. Our results indicate that a four-quark structure is 
acceptable for the resonances observed by BELLE and BABAR: $D_0^{0}(2308)$ 
and $D_{sJ}^{+}(2317)$ respectively, but not for the resonances observed 
by FOCUS: $D_0^{0,+}(2405)$.  
\end{abstract}

\maketitle

%\section{Introduction}
Recently the first observations of the scalar charmed mesons have been 
reported. The very narrow $D_{sJ}^+(2317)$ was first discovered in the
$D_s^+\pi^0$ channel by the BABAR Collaboration \cite{babar} and its
existence was confirmed by CLEO \cite{cleo}, BELLE \cite{belle1} and
FOCUS \cite{focus} Collaborations. Its mass was commonly measured as
$2317 \MeV$, which is approximately $160 \MeV$ below the prediction of 
the very successful quark model for the charmed mesons \cite{god}. 
The BELLE Collaboration \cite{belle2} has also reported the observation of 
a rather broad scalar meson $D_0^{0}(2308)$, and the FOCUS Collaboration
\cite{focus2} reported evidence for broad structures in both neutral and
charged final states that, if interpreted as resonances in the $J^P=0^+$
channel, would be  the $D_0^{0}(2407)$ and the $D_0^{+}(2403)$ mesons. 
While the 
mass of the scalar meson, $D_0^{0}(2308)$, observed by  BELLE Collaboration
is also bellow the prediction of ref.~\cite{god} (approximately $100 \MeV$),
the masses of the states observed by FOCUS Collaboration are in complete
agreement with ref.~\cite{god}.

Due to its low mass, the structure of the meson $D_{sJ}^+(2317)$ has been 
extensively debated. It has been interpreted as 
a $c\bar{s}$ state \cite{dhlz,bali,ukqcd,ht,nari}, two-meson molecular 
state \cite{bcl,szc}, $D-K$- mixing \cite{br},  
four-quark states \cite{ch,tera,mppr} or a mixture between two-meson
and four-quark states \cite{bpp}. The same analyses would also apply
to the meson $D_0^{0}(2308)$.

In the light sector the idea that the
scalar mesons could be four-quark bound states is not new \cite{jaffe} and,
therefore, it is natural to consider analogous states in the charm sector.

We propose that the resonances observed by BELLE \cite{belle2} and FOCUS
\cite{focus2} Collaborations be considered as two different resonances.
In this work we use the  method of QCD  sum rules (QCDSR) \cite{svz}
to study the two-point functions of the scalar mesons, $D_{sJ}(2317)$, 
$D_0(2308)$ and $D_0(2405)$ considered as four-quark states. 
The use of the QCD sum rules to study the charmed scalar mesons
was already done in refs.~\cite{dhlz,ht,nari}, but
in these calculations they were interpreted as two-quark states.

In a recent calculation \cite{sca} some of us have considered that the 
lowest lying scalar 
mesons are $S$-wave bound states of a diquark-antidiquark pair. As 
suggested in ref.~\cite{jawil} the diquark was taken to be a spin zero 
colour anti-triplet. We extend this prescription
to the charm sector and, therefore, the corresponding interpolating fields 
containing zero, one and two strange quarks are:
\beqa
j_0&=&\epsilon_{abc}\epsilon_{dec}(q_a^TC\gamma_5c_b)
(\bar{u}_d\gamma_5C\bar{d}_e^T),
\nn\\
j_s&=&{\epsilon_{abc}\epsilon_{dec}\over\sqrt{2}}\left[(u_a^TC
\gamma_5c_b)(\bar{u}_d\gamma_5C\bar{s}_e^T)+u\leftrightarrow d\right],
\nn\\
j_{ss}&=&\epsilon_{abc}\epsilon_{dec}(s_a^TC
\gamma_5c_b)(\bar{q}_d\gamma_5C\bar{s}_e^T),
\label{int}
\enqa
where $a,~b,~c,~...$ are colour indices, $C$ is the charge conjugation
matrix and 
$q$ represents the quark $u$ or $d$ according to the charge of the meson.
Since $D_{sJ}$ has one $\bar{s}$ quark, we choose the $j_s$ current to 
have the same quantum numbers of $D_{sJ}$, which is supposed to be
an isoscalar. However, since we are working in the SU(2) limit, the
isoscalar and isovector states are mass degenerate and, therefore, this
particular choice has no relevance here.

The QCDSR for the charmed scalar mesons are constructed from the two-point
correlation function
\beq
\Pi(q)=i\int d^4x ~e^{iq.x}\lag 0 |T[j_S(x)j^\dagger_S(0)]|0\rag.
\lb{2po}
\enq

The coupling of the scalar meson, $S$, to the scalar current, $j_S$,  can be
parametrized in terms of the meson decay constant $f_S$ as \cite{sca}:
$\lag 0 | j_S|S\rag =\sqrt{2}f_Sm_S^4$,
therefore, the phenomenological side of Eq.~(\ref{2po}) can be written as
\beq
\Pi^{phen}(q^2)={2f_S^2m_S^8\over m_S^2-q^2}+\cdots\;,
\lb{phe}
\enq
where the dots denote higher resonance contributions that will be 
parametrized, as usual, through the introduction of the continuum threshold
parameter $s_0$ \cite{io1}.

In the OPE side we work at leading order and consider condensates up to 
dimension six. We deal with the strange quark as a light one and consider
the diagrams up to order $m_s$. To keep the charm quark mass finite, we
use the momentum-space expression for the charm quark propagator. We follow 
ref.~\cite{su} and calculate the light quark part of the correlation
function in the coordinate-space, which is then Fourier transformed to the
momentum space in $D$ dimensions. The resulting light-quark part is combined 
with the charm-quark part before it is dimensionally regularized at $D=4$.

We can write the correlation function in the OPE side in terms of a 
dispersion relation:
\beq
\Pi^{OPE}(q^2)=\int_{m_c^2}^\infty ds {\rho(s)\over s-q^2}\;,
\lb{ope}
\enq
where the spectral density is given by the imaginary part of the correlation
function: $\rho(s)={1\over\pi}\mbox{Im}[\Pi^{OPE}(s)]$. After making a Borel
transform on both sides, and transferring the continuum contribution to
the OPE side, the sum rule for the scalar meson $S$ can be written as
\beq
2f_S^2m_S^8e^{-m_S^2/M^2}=\int_{m_c^2}^{s_0}ds~ e^{-s/M^2}~\rho_S(s)\;,
\lb{sr}
\enq
where $\rho_S(s)=\rho^{pert}(s)+\rh^{m_s}(s)+\rh^{\qq}(s)+\rh^{\lag G^2\rag}
(s)+\rh^{mix}(s)+\rh^{\qq^2}(s)+\rh^{\lag G^3\rag}(s)$, with
\beq
\rho^{pert}(s)={1\over 2^{10} 3\pi^6}\int_\La^1 d\al\left({1-\al\over\al}
\right)^3(m_c^2-s\al)^4,
\enq
\beqa
\rho^{\lag G^2\rag}(s)&=&{\Gd\over 2^{10}\pi^6}\int_\La^1 d\al~(m_c^2-s\al)
\left[{m_c^2\over9}\left({1-\al\over\al}\right)^3+\right.
\nn\\&+&
\left.(m_c^2-s\al)\left({1-\al\over2\al}+{(1-\al)^2\over4\al^2}\right)
\right],
\enqa
\beq
\rho^{\lag G^3\rag}(s)={\G3\over 2^{12} 9\pi^6}\int_\La^1 d\al\left({1-\al
\over\al}\right)^3(3m_c^2-s\al),
\enq
which are common to all three resonances and where the lower limit of the 
integrations is given by $\La=m_c^2/s$. From $j_0$ we get: $\rh^{m_s}(s)=0$,
\beq
\rho^{\qq}(s)=-{m_c\qq\over 2^{6}\pi^4}\int_\La^1 d\al\left({1-\al\over\al}
\right)^2(m_c^2-s\al)^2,
\enq
\beqa
\rho^{mix}(s)&=&{m_c\mix\over 2^{6}\pi^4}\bigg[{1\over2}\int_\La^1 d\al
\left({1-\al\over\al}\right)^2(m_c^2-s\al)+
\nn\\
&-&\int_\La^1 d\al{1-\al\over\al}(m_c^2-s\al)\bigg],
\enqa
\beq
\rho^{\qq^2}(s)=-{\qq^2\over 12\pi^2}\int_\La^1 d\al~(m_c^2-s\al).
\enq
From $j_{s}$ we get: $\rh^{m_s}(s)=0$,
\beqa
\rho^{\qq}(s)={1\over 2^{6}\pi^4}\int_\La^1 d\al~{1-\al\over\al}
(m_c^2-s\al)^2
\bigg[
\nn\\
-\qq\left(2m_s+m_c{1-\al\over\al}\right)+m_s\ss\bigg],
\enqa
\beqa
\rho^{mix}(s)&=&{1\over 2^{6}\pi^4}\int_\La^1 d\al~(m_c^2-s\al)\bigg[-{m_s\mixs
\over6}
\nn\\
&+&\mix\bigg(-m_s(1-\ln(1-\al))
\nn\\
&-&m_c{1-\al\over\al}\left(1-{1-\al\over2\al}\right)\bigg)
\bigg]
\enqa
\beq
\rho^{\qq^2}(s)=-{\qq\ss\over 12\pi^2}\int_\La^1 d\al~(m_c^2-s\al).
\enq
Finally from $j_{ss}$ we get
\beq
\rh^{m_s}(s)=-{m_sm_c\over 2^{8} 3\pi^6}\int_\La^1 d\al\left({1-\al\over\al}
\right)^3(m_c^2-s\al)^3,
\enq
\beqa
\rho^{\qq}(s)={1\over 2^{6}\pi^4}\int_\La^1 d\al{1-\al\over\al}(m_c^2-s\al)^2
\bigg[
\nn\\
\ss\left(2m_s-m_c{1-\al\over\al}\right)-2m_s\qq\bigg],
\enqa
\beqa
\rho^{mix}(s)&=&{1\over 2^{6}\pi^4}\int_\La^1 d\al~(m_c^2-s\al)\bigg[{\mixs
\over2}\bigg({m_s\over3}
\nn\\
&-&m_s{1-\al\over\al}
-m_c{1-\al\over\al}\left(1-{1-\al\over2\al}\right)\bigg)
\nn\\
&-&m_s\mix(1-\ln(1-\al))\bigg],
\enqa
\beq
\rho^{\qq^2}(s)=-{\qq\ss\over 12\pi^2}\int_\La^1 d\al~(m_c^2-s\al).
\enq
For the charm quark propagator with two and three gluons attached we use
the momentum-space expressions given in ref.~\cite{rry}.

In order to get rid of the meson decay constant and
extract the resonance mass, $m_S$, we first take the derivative
of Eq.~(\ref{sr}) with respect to $1/M^2$ and then we divide it by
Eq.~(\ref{sr}) to get
\beq
m_S^2={\int_{m_c^2}^{s_0}ds ~e^{-s/M^2}~s~\rh_S(s)\over\int_{m_c^2}^{s_0}ds 
~e^{-s/M^2}~\rh_S(s)}\;.
\lb{m2}
\enq

In the numerical analysis of the sum rules, the values used for the quark
masses and condensates are: $m_s=0.13\,\GeV$, $m_c=1.2\,\GeV$, 
$\lag\bar{q}q\rag=\,-(0.23)^3\,\GeV^3$,
$\langle\overline{s}s\rangle\,=0.8\lag\bar{q}q\rag$,  
$\lag\bar{q}g\si.Gq\rag=m_0^2
\lag\bar{q}q\rag$ with $m_0^2=0.8\,\GeV^2$, $\lag g^2G^2\rag=0.5~\GeV^4$
and $\lag g^3G^3\rag=0.045~\GeV^6$. The value for the quark condensate was
obtained using the Gell-Mann - Oakes - Renner relation, and the mass of the
light quarks, $m_u+m_d=14\MeV$, at the renormalization scale of $1\GeV$ 
\cite{gale}. Since the charm quark mass introduces a natural scale in the 
problem, we chose to work at the renormalization scale of $m_c\sim1\GeV$.

\begin{figure}[h] \label{fig1}
%\leavevmode
\centerline{\epsfig{figure=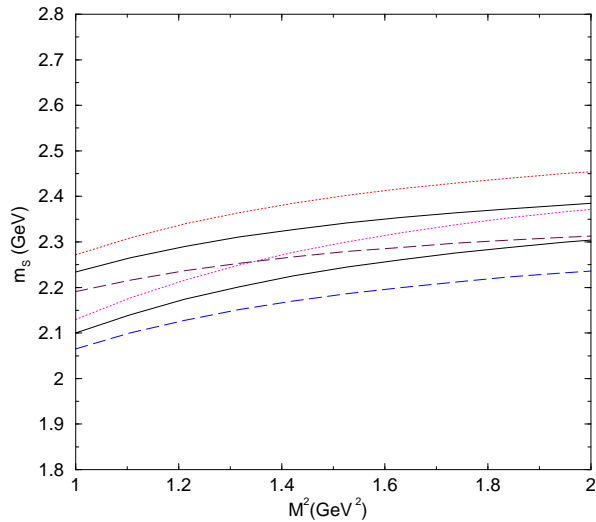,height=70mm}}
%,width=70mm,angle=0}}
\caption{The $D_0^{(0s)}$ mass (the lower dashed, solid and dotted lines)
and the $D_0^{(1s)}$ mass (the upper dashed, solid and dotted lines),
as a function of the Borel mass for different
values of the continuum threshold.  Dashed lines: $\sqrt{s_0}=2.6\GeV$; solid
lines: $\sqrt{s_0}=2.7\GeV$; dotted lines: $\sqrt{s_0}=2.8\GeV$.} 
\end{figure} 

We call $D_0^{(0s)}$, $D_{0}^{(1s)}$ and $D_{0}^{(2s)}$ the scalar charmed
 mesons represented by $j_0$,
$j_s$ and $j_{ss}$ (in Eq.~(\ref{int})) respectively.
In Figs. 1 and 2 we show the masses of these three resonances
as a function of the Borel mass for different values of the continuum 
threshold.

The Borel window was fixed in such way that the pole contribution is always
between 80\% and 20\% of the total contribution.

\begin{figure}[h] \label{fig2}
%\leavevmode
\centerline{\epsfig{figure=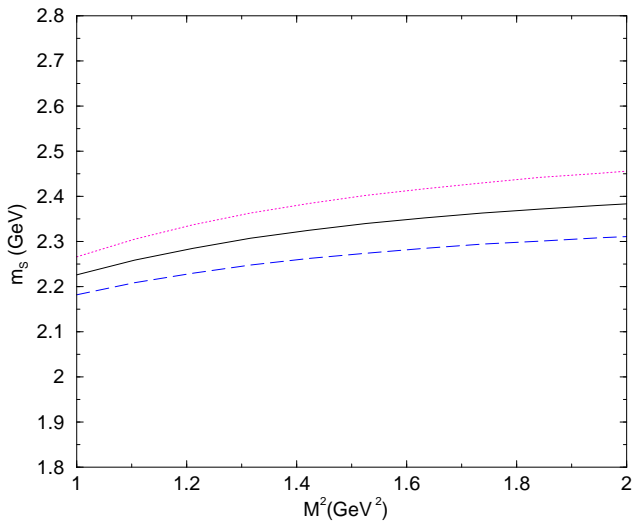,height=70mm}}
%,width=70mm,angle=0}}
\caption{The $D_{0}^{(2s)}$ mass as a function of the Borel mass 
for different
values of the continuum threshold.  Dashed line: $\sqrt{s_0}=2.6\GeV$; solid
line: $\sqrt{s_0}=2.7\GeV$; dotted line: $\sqrt{s_0}=2.8\GeV$.}
\end{figure} 

Fixing $\sqrt{s_0}=2.7\GeV$ and varying the charm quark and the strange
quark masses in the intervals: $1.1\leq m_c\leq 1.3\GeV$ and 
$0.11\leq m_s\leq 0.15\GeV$, we get results for the resonance masses still
between the lower and upper lines in figures 1 and 2. A bigger value
for the charm quark mass makes the results more stable as a function of
the Borel mass. One can also vary the value of the quark condensate. Keeping 
the continuum threshold and the quark masses fixed at $\sqrt{s_0}=2.7\GeV$,
$m_c= 1.2\GeV$ and $m_s=0.13\GeV$ and varying the quark condensate
in the interval: $\lag\bar{q}q\rag=\,(-0.23\pm0.01\GeV)^3$, we get a bigger
(smaller) result for the resonance masses using a smaller (bigger) value of 
the condensate.  
In Fig.~3 we show the
the mass of the $D_{0}^{(1s)}$ state, as a function of the Borel mass, for 
the combination of the values of the continuum threshold and quark 
condensate that gives the lower and upper limits for the $D_{0}^{(1s)}$ mass.

\begin{figure}[h] \label{fig3}
%\leavevmode
\centerline{\epsfig{figure=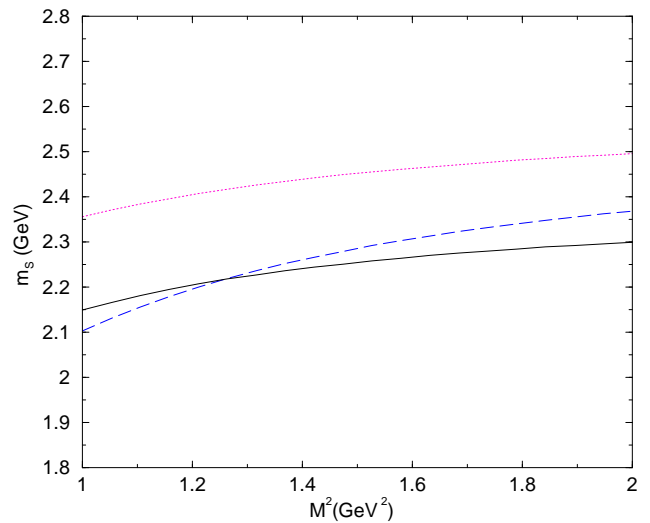,height=70mm}}
%,width=70mm,angle=0}}
\caption{The $D_{0}^{(1s)}$ mass as a function of the Borel mass 
for different values of the continuum threshold and quark condensate.  Solid 
line: $\sqrt{s_0}=2.6\GeV$ and $\lag\bar{q}q\rag(1\GeV)=\,(-0.24\GeV)^3$; 
dotted line: $\sqrt{s_0}=2.8\GeV$ and $\lag\bar{q}q\rag(1\GeV)=\,(-0.22\GeV)
^3$; dashed line: $\sqrt{s_0}=2.6\GeV$, $\lag\bar{q}q\rag(2\GeV)=\,
(-0.267\GeV)^3$
and $m_s(2\GeV)=0.10\GeV$.}
\end{figure} 

In ref.\cite{jala} it was shown that the renormalization scale was an important
source of uncertainty, in the analysis of the $B$ meson decay constant. To 
check how the change of the scale would change our results we also show, 
through the dashed line in Fig.~3 , the result for the $D_{0}^{(1s)}$ 
resonance 
mass using the values of the strange quark mass and quark condensate at the 
scale $2\GeV$: $\lag\bar{q}q\rag(2\GeV)=\,(-0.267\GeV)^3$ and 
$m_s(2\GeV)=0.10\GeV$ \cite{jala}. We see that we get a less stable result for
 the  ressonance mass, but it is still compatible with the results at the 
scale $1\GeV$, considering the variation in the continuum threshold. Therefore,
 we conclude that it is the variation of the continuum threshold that causes 
the most significant variations in the resonance masses, and it is our most 
important source of uncertaintiy.

Comparing figures 1 and 2 we see that 
the $D_{0}^{(1s)}$ and $D_{0}^{(2s)}$
resonance masses  are basicaly degenerated, while the
mass of $D_{0}^{(0s)}$ is around $100\MeV$ smaller than the others. While
it is natural to expect that the inclusion of a strange quark would increase
the resonance mass by around the strange quark mass (as was the case when 
one goes from $D_{0}^{(0s)}$ to $D_{0}^{(1s)}$), it is really interesting
to observe that this does not happen when one goes from $D_{0}^{(1s)}$
to $D_{0}^{(2s)}$. In terms of the OPE contributions, we can trace this
behavior to the fact that the quark condensate term is smaller in
$D_{0}^{(2s)}$ than in $D_{0}^{(1s)}$ (due to the change from $m_c\qq$ to 
$m_c\ss$), however the inclusion of the term
proportional to $m_sm_c$ (which is not present in $D_{0}^{(1s)}$), 
compensates this decrease.

Considering the variations on the quark masses, the quark condensate and on 
the continuum 
threshold discussed above, in the Borel window considered here our results 
for the ressonance masses are given in Table I.
\begin{center}
\small{{\bf Table I:} Numerical results for the resonance masses}
\\
\vskip0.3cm
\begin{tabular}{|c|c|c|c|}  \hline
resonance & $D_{0}^{(0s)}$  & $D_{0}^{(1s)}$ & $D_{0}^{(2s)}$   \\
\hline
mass (GeV) & $~2.22\pm0.21~$ &$~2.32\pm0.18~$  & $~2.30\pm0.20~$\\
\hline
\end{tabular}\end{center}

Comparing the results in Table I with the resonance masses given by
BABAR, BELLE and FOCUS: $D_{sJ}^+(2317)$,
$D_{0}^0(2308)$ and $D_{0}^{0,+}(2405)$, we see that we can 
identify the four-quark states represented by $D_{0}^{(1s)}$ and 
$D_{0}^{(2s)}$ with the BABAR and BELLE resonances respectively. However,
we do not find a four-quark state whose mass is compatible with the
FOCUS resonances, $D_{0}^{0,+}(2405)$. Therefore, we associate the
FOCUS resonances, $D_{0}^{0,+}(2405)$,
with a scalar $c\bar{q}$ state, since its mass is completly in agreement
with the predictions of the quark model in ref.~\cite{god}. It is also 
interesting to point out that a mass of about $2.4~\GeV$ is also compatible
with the the QCD sum rule calculation for a $c\bar{q}$ scalar meson 
\cite{ht}.

One can still argue that while a pole approximation is justified for
the very narrow BABAR resonance, this may not be the case for the
rather broad BELLE and FOCUS resonances. To check if the width of the
resonances could modify the pattern observed in the masses of the
four-quark states, we have modified the phenomenological side of the
sum rule, in Eq.~(\ref{sr}), through the introduction of a 
Breit-Wigner-type resonance form:
\beq
\Pi^{phen}(M^2)=2f_S^2m_S^8\int_{(m_\pi+m_D)^2}^{s_0} ds~e^{-s/M^2}\rho_{BW}
(s)\;,
\lb{phenbw}
\enq
where
\beq
\rho_{BW}(s)={1\over\pi}{\Gamma(s)m_S\over(s-m_S^2)^2+m_S^2\Gamma(s)^2},
\lb{bw}
\enq
with $\Gamma(s)=\Gamma_0{\sqrt{\lambda(s,m_D^2,m_\pi^2)\over\lambda(m_S,
m_D^2,m_\pi^2)}}{m_S^2\over s}$,
and $\lambda(x,y,z)=x^2+y^2+z^2-2xy-2xz-2yz$.

\begin{figure}[h] \label{fig4}
%\leavevmode
\centerline{\epsfig{figure=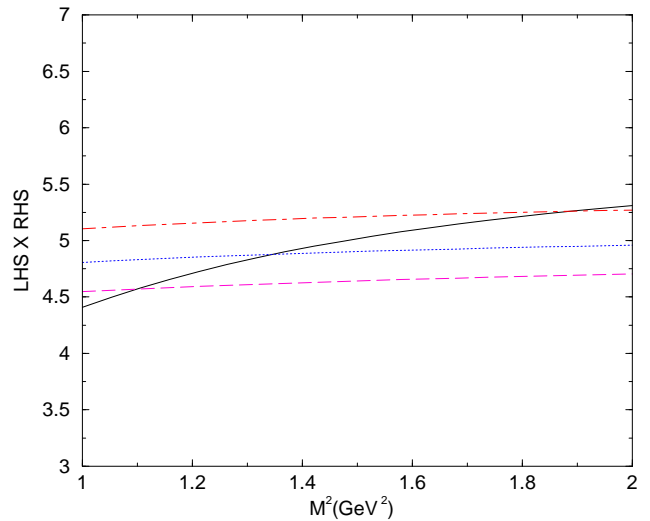,height=70mm}}
%,width=70mm,angle=0}}
\caption{The RHS (solid line) and the LHS of the sum rule in Eq.~(\ref{m2bw})
for $D_{0}^{(0s)}$, for different values of the resonance mass. 
Dashed line: $m_S=2.1\GeV$; 
dotted line: $m_S=2.2\GeV$; dot-dashed line:  $m_S=2.3\GeV$.} 
\end{figure} 
Of course now we can not 
obtain an expression for the resonance mass as Eq.~(\ref{m2}). However,
we can still use the resonance mass as a parameter to compare the 
compatibility between the right-hand side (RHS) and the left-hand side
(LHS) of the sum rule in Eq.~(\ref{m2bw}):
\beq
{\int_{(m_\pi+m_D)^2}^{s_0}ds e^{-s/M^2}s\rh_{BW}(s)\over\int_{(m_\pi+m_D
)^2}^{s_0}ds e^{-s/M^2}\rh_{BW}(s)}={\int_{m_c^2}^{s_0}ds e^{-s/M^2}s
\rh_S(s)\over\int_{m_c^2}^{s_0}ds e^{-s/M^2}\rh_S(s)}\;.
\lb{m2bw}
\enq

In Fig.4 we show the RHS (solid line) and the LHS of  Eq.~(\ref{m2bw})
for $D_{0}^{(0s)}$,
for three different values of the resonance mass, with $\Gamma_0=280~\MeV$
and $\sqrt{s_0}=
2.7~\GeV$. We see that the best agreement is obtained for $m_S\sim2.2~\GeV$,
which shows that the inclusion of the width does not change the value
of the mass obtained for the resonance.

We have presented a QCD sum rule study of the charmed scalar mesons 
considered as diquark-antidiquark states. We found that the masses
of the BABAR,  $D_{sJ}^+(2317)$, and BELLE, $D_{0}^0(2308)$, resonances
can be reproduced by the four-quark states $(cq)(\bar{q}\bar{s})$
and $(cs)(\bar{u}\bar{s})$ respectively. However, the mass of the FOCUS
resonance, $D_{0}^{0,+}(2405)$, which we believe is not the same measured by
BELLE, can not be reproduced in the four-quark state
picture considered here. Therefore, we interpret it as a  normal
$c\bar{q}$ state, since its mass is in complete  agreement
with the predictions of the quark model in ref.~\cite{god}. We also obtain 
a mass of $\sim 2.2~\GeV$ for
a four-quark scalar state $(cq)(\bar{u}\bar{d})$ which was not yet
observed, and that should be also rather broad.

\vspace{1cm}
 
\underline{Acknowledgements}: 
We would like to thank I. Bediaga  for fruitful discussions. 
This work has been supported by CNPq and FAPESP.

\end{document}